\documentclass[a4paper,11pt]{article}
\usepackage{pos}
\usepackage{amsmath}
\usepackage{calrsfs}
\usepackage{autobreak}

\newcommand{\bracket}[1]{( #1 )}
\DeclareMathOperator{\Ti2}{Ti_2}

\title{NNLO corrections to SIDIS coefficient functions}
\author*[a]{Leonardo Bonino}
\author[a]{Thomas Gehrmann}
\author[a]{Markus L\"ochner}
\author[a]{Kay Sch\"onwald}
\author[b]{Giovanni Stagnitto}

\affiliation[a]{Physik-Institut, Universit\"at Z\"urich,Winterthurerstrasse 190, 8057 Z\"urich, Switzerland
}

\affiliation[b]{Universit\`{a} degli Studi di Milano-Bicocca \& INFN, Piazza della Scienza 3, 20216 Milano, Italy}

\emailAdd{leonardo.bonino@physik.uzh.ch}

\abstract{Hadron production in lepton-proton scattering (semi-inclusive deep inelastic scattering, SIDIS) probes the structure of hadrons at a higher level of detail than fully inclusive processes. A wealth of SIDIS data is available especially from fixed-target experiments. Here we review our calculation for the NNLO corrections to the full set of polarized and unpolarized SIDIS coefficient functions and present some selected analytical expressions. Our results enable for the first time a fully consistent treatment of hadron fragmentation processes in polarized and unpolarized DIS at NNLO and provide the basis for studies of hadron structure, hadron fragmentation and identified particle cross sections at colliders.}

\FullConference{Loops and Legs in Quantum Field Theory (LL2024)\\
 14-19, April, 2024\\
Wittenberg, Germany\\}

\newcommand{\der}{\mathrm{d}}

\newcommand{\Fcal}{\pazocal{F}}
\newcommand{\Ccal}{\pazocal{C}}
\newcommand{\DC}{\Delta C}
\newcommand{\hx}{\hat{x}}
\newcommand{\hz}{\hat{z}}
\newcommand{\qb}{\bar{q}}
\newcommand{\qp}{q^{\prime}}
\newcommand{\qbp}{\bar{q}^{\prime}}

\newcommand\MSbar{\overline{\mathrm{MS}}}

\newcommand{\HVBM}{HVBM}

\DeclareMathAlphabet{\pazocal}{OMS}{zplm}{m}{n}

\begin{document}
\maketitle

\section{Introduction}

The production of various hadron species in polarized and unpolarized semi-inclusive deep-inelastic lepton-nucleon scattering
(SIDIS) has been measured extensively~\cite{EuropeanMuon:1991sne,ZEUS:1995acw,H1:1996muf,HERMES:2012uyd,COMPASS:2016xvm,COMPASS:2010hwr,HERMES:2006jyl}.  By considering different flavour combinations of the incoming and outgoing (identified) parton one can probe the detailed quark and antiquark flavour decomposition of parton distribution functions (PDFs).  In particular, due to the limited data from hadron-hadron collisions,  polarized SIDIS measurements play an important role in the determination of the flavor structure spin-dependent PDFs~\cite{deFlorian:2009vb}, which encode the proton’s spin structure. Polarized PDFs describe the probability of finding a parton with a given momentum fraction $x$ at a resolution 
scale $Q^2$ with its helicity aligned or anti-aligned to the nucleon’s spin. With $ f(x,Q^2)$ being the well-established (unpolarized) PDFs we can write: 
\begin{align}
 f(x,Q^2) =& f^+(x,Q^2)+f^-(x,Q^2)\,, \nonumber\\
 \Delta f(x,Q^2)=&  f^+(x,Q^2)-f^-(x,Q^2)\,,
\end{align}
where $^\pm$ refers to the relative orientation of the parton helicity with respect to the parent nucleon spin.
Polarized PDFs $\Delta f(x,Q^2)$ are accessible in spin asymmetries where both probe and target are longitudinally polarized.

Up to now,  global studies on $\Delta f(x,Q^2)$~\cite{deFlorian:2009vb,Bertone:2024taw} could only be performed in a self-consistent manner up to NLO since higher-order corrections to the SIDIS coefficient functions have not been available.  

Our recent results \cite{Bonino:2024qbh,Bonino:2024wgg} for the full set of NNLO QCD corrections to the polarized and unpolarized SIDIS coefficient functions will enable consistent NNLO global fits to observables with identified hadrons for the first time.  After briefly recalling the kinematics of SIDIS, and the method employed in our calculation, we provide selected analytical expressions for the pure-singlet channels.  Numerical studies and comparisons to data are presented in \cite{Bonino:2024qbh,Bonino:2024wgg}.

%----------------------------------------
\section{Kinematics of SIDIS}
%----------------------------------------

We consider the production of an unpolarized hadron $h$ (e.g\@. a pion) from the scattering of a lepton off a nucleon.  Both leptons and nucleon can be longitudinally polarized.  We describe (polarized) semi-inclusive deep-inelastic scattering as
$\overset{\scriptscriptstyle(\to)}{\ell\phantom{.}}(k)\,\overset{\scriptscriptstyle(\to)}{p\phantom{.}}(P)\to\ell(k^{\prime})\,h(P_h)\,X$, with some inclusive final-state
radiation $X$, following the notation of~\cite{deFlorian:1997zj,Anderle:2013lka}. For a polarized scattering we require both lepton and nucleon to be longitudinally polarized.  With $q=k-k^{\prime}$ we denote the momentum transfer
between the leptonic and hadronic systems, and with $y=(P\cdot
q)/(P\cdot k)$ the associated energy transfer at 
virtuality $Q^2=-q^2$. The quantities
\begin{align}
x=&\frac{Q^2}{2P\cdot q} & \rm{and}& &  z=&\frac{P\cdot P_h}{P\cdot q}
\end{align}
are the momentum fractions of the nucleon carried by the incoming parton
($x$), and of the outgoing parton carried by the identified
hadron ($z$) at Born level. The variables $x$, $y$ and $Q^2$ are not independent and are related by $s=Q^2/(xy)$, where $\sqrt{s}$ is the lepton-nucleon system center-of-mass energy.

For $Q\ll M_Z$ only (highly) virtual photons are exchanged,
and the spin-averaged triple-differential cross section reads
\begin{equation}\label{d3sigdxdydz}
  \frac{\der^3\sigma^h}{\der x \der y \der z} = \frac{4\pi\alpha^2}{Q^2} \bigg[ \frac{1+(1-y)^2}{2y} \Fcal^h_T(x,z,Q^2)
    + \frac{1-y}{y} \Fcal^h_L(x,z,Q^2) \bigg] \, ,
\end{equation}
where $e^2=4\pi\alpha$. 

The transverse $\Fcal^h_T$ and longitudinal $\Fcal^h_L$ unpolarized SIDIS structure
functions are given by the sum over all partonic channels of the convolution
between the PDF $f_p$ for a parton $p$, the fragmentation function (FF) $D^h_{p^{\prime}}$ of a parton $p^{\prime}$ into
the hadron $h$,  and the coefficient function $\Ccal^i_{p' p}$ for the
transition $p\to p^{\prime}$,
\begin{align}
  \Fcal_i^h(x,z,Q^2) = \sum_{p,p'} \int_x^1 \frac{\der\hx}{\hx}
  \int_z^1 \frac{\der\hz}{\hz} f_p\left(\frac{x}{\hx},\mu_F^2\right)
  D_{p'}^h\left(\frac{z}{\hz},\mu_A^2\right)\Ccal^i_{p' p}\left(\hx,\hz,Q^2,\mu_R^2,\mu_F^2,\mu_A^2\right)\,, 
\end{align}
with $i = T,L$.
In the above expression $\mu_F$ and $\mu_A$ are the initial and final-state factorization scales respectively, and $\mu_R$ is the renormalization scale.  The differential cross-section of eq.~\eqref{d3sigdxdydz} is sometimes written in terms of the structure functions $F^h_1$ and $F^h_2$ that are related to the transverse and longitudinal ones by  $\Fcal^h_T=2F^h_1$ and $\Fcal^h_L=(F^h_2/x-2F^h_1)$.

For longitudinally polarized lepton and nucleon, the (polarized $\Delta$) differential cross section is given by \cite{deFlorian:1997zj}
\begin{equation}\label{d3Dsigdxdydz}
  \frac{\der^3\Delta\sigma^h}{\der x \der y \der z} = \frac{4\pi\alpha^2}{Q^2} (2-y) g_1^h(x,z,Q^2) \,  .
\end{equation}

The polarized SIDIS structure function $g^h_1$ is given by the sum over all partonic channels of the convolution
between the polarized PDF $\Delta f_p$ for a parton $p$, the FF $D^h_{p^{\prime}}$ of parton $p^{\prime}$ into hadron $h$, and the polarized coefficient function $\Delta \Ccal_{p' p}$ for the partonic
transition $p\to p^{\prime}$:
\begin{equation}\label{eq:g1h}
  2g^h_1(x,z,Q^2)= \sum_{p,p'} \int_x^1 \frac{\der\hx}{\hx}
  \int_z^1 \frac{\der\hz}{\hz} \Delta f_p\bigl(\frac{x}{\hx},\mu_F^2\bigr) D_{p'}^h\bigl(\frac{z}{\hz},\mu_A^2\bigr)  
 \Delta \Ccal_{p' p}\bigl(\hx,\hz,Q^2,\mu_R^2,\mu_F^2,\mu_A^2\bigr).
\end{equation}

 The polarized and unpolarized SIDIS coefficient
functions $\mathcal{C}_{p'p}\in\{\Ccal^{T}_{p'p},\Ccal^{L}_{p'p},\Delta\Ccal_{p'p}\}$ encode the hard-scattering part of the process, and can be computed in
perturbative QCD. Their perturbative expansion in the strong coupling constant $\alpha_s$ reads
\begin{equation}
  \mathcal{C}_{p' p} = \mathcal{C}^{(0)}_{p' p}
  + \frac{\alpha_s(\mu_R^2)}{2\pi}  \mathcal{C}^{(1)}_{p' p}+ \biggl(\frac{\alpha_s(\mu_R^2)}{2\pi}\biggr)^2   \mathcal{C}^{(2)}_{p' p}
  + \pazocal{O}(\alpha_s^3)\,  , 
\end{equation} 
where the NLO corrections can for example be found in~\cite{deFlorian:1997zj}.

In \cite{Bonino:2024qbh,Bonino:2024wgg} we presented results for the NNLO corrections
$\mathcal{C}^{(2)}_{p' p}$ to all partonic channels appearing at this order. The seven partonic channels
appearing at $\pazocal{O}(\alpha_s^2)$ are, following the notation of~\cite{Anderle:2016kwa,Bonino:2024qbh}:
\begin{align}\label{CFNNLOlist}
\mathcal{C}^{(2)}_{qq}&=e_q^2 \mathcal{C}^{\mathrm{NS}}_{qq}+\biggl( \sum_j e^2_{q_j}\biggr)\mathcal{C}^{\mathrm{PS}}_{qq} \, , \nonumber \\
\mathcal{C}^{(2)}_{\qb q}&=e_q^2\mathcal{C}_{\qb q} \, , \nonumber \\
\mathcal{C}^{(2)}_{\qp q}&=e_q^2 \mathcal{C}^{1}_{\qp q}+e_{\qp}^2 \mathcal{C}^{2}_{\qp q}+e_q e_{\qp}\mathcal{C}^{3}_{\qp q} \, , \nonumber \\
\mathcal{C}^{(2)}_{\qbp q}&=e_q^2\mathcal{C}^{1}_{\qp q}+e_{\qp}^2 \mathcal{C}^{2}_{\qp q}-e_q e_{\qp} \mathcal{C}^{3}_{\qp q} \, , \nonumber \\
\mathcal{C}^{(2)}_{gq}&=e_q^2 \mathcal{C}_{gq} \, , \nonumber \\
\mathcal{C}^{(2)}_{qg}&=e_q^2 \mathcal{C}_{qg} \, , \nonumber \\
\mathcal{C}^{(2)}_{gg}&=\biggl( \sum_j e_{q_j}^2 \biggr) \mathcal{C}_{gg} \,  .
\end{align}
With $\overset{\textbf{\fontsize{5pt}{5pt}\selectfont(--)}\prime}{q\phantom{.}}$ we indicate an (anti-)quark of flavor different from $q$ and the sums run over all quarks flavors.  The NS and PS superscripts in the quark-to-quark channel denote the non-singlet and the pure-singlet components respectively.

In Section \ref{sec:res} we present analytical results for the coefficients $C^{T,\mathrm{PS}}_{qq}$ and $\DC^{\mathrm{PS}}_{qq}$.
 
\section{Method}
The coefficient functions are computed by applying suitable projectors from the respective parton-level
subprocess matrix elements with incoming kinematics fixed by $Q^2$ and
$\hat{x}$, which are then integrated over the final-state phase space.  In the final-state integration the momentum fraction of the parton $p'$ is fixed to $\hat{z}$ and the remaining extra radiation $X$ is fully integrated.

At NNLO in QCD, three types of parton-level contributions must be taken into
account, relative to the underlying Born-level process: two-loop virtual
corrections (double-virtual, VV), one-loop corrections to single real radiation
processes (real-virtual, RV) and tree-level double real radiation processes
(RR). 

The one-loop squared matrix elements (RV contributions) can be expressed in terms of one-loop bubble and box integrals.  These integrals are known in exact form in $d = 4-2\epsilon$, where $d$ denotes the dimension of space-time.
The associated phase space integral is fully constrained for fixed $\hat{x}$ and $\hat{z}$ and only expansions in the end-point
distributions in $\hat{x}=1$ and $\hat{z}=1$ are required.  For a generic RV contribution with kinematics $k_i+q\rightarrow k_j^{\mathrm{id.}}+k_k$ we can thus write
\begin{align}
\tilde{\mathcal{C}}^{RV}_{p^{\prime}p}(\hx,\hz,Q^2,\epsilon)&\propto\int \der \Phi_2(k_j,k_k;k_i,q)\delta \left(\hz-\hx\frac{(k_i+k_j)^2}{Q^2}\right)\left|\pazocal{M}^{RV}_{p^\prime p}\right|^2 \propto \pazocal{J}(\hx,
\hz)\left|\pazocal{M}^{RV}_{p^\prime p}\right|^2(\hx,\hz)\,  ,
\end{align}
where with $\,\tilde{}\,$ we indicate a bare quantity. The Jacobian factor is given by
\begin{align}
\pazocal{J}(\hx,\hz)=(1-\hx)^{-\epsilon}\hx^{\epsilon}\hz^{-\epsilon}(1-\hz)^{-\epsilon}\, .
\end{align}
To avoid ambiguities associated with the analytic
continuation of the one-loop master integrals, we segment the parameter space of the RV
contribution	into four sectors: $(\hat{x}\leq 0.5, \hat{x}\leq
\hat{z}\leq 1-\hat{x})$, $(\hat{z}\leq 0.5, \hat{z}<\hat{x}\leq 1-\hat{z})$,
$(\hat{x}>0.5,1-\hat{x}< \hat{z}<\hat{x})$ and
$(\hat{z}>0.5,1-\hat{z}<\hat{x}\leq \hat{z})$, where the contributions are manifestly real~\cite{Gehrmann:2022cih}. We checked the continuity of our expressions across the boundaries of the regions.

The RR contributions are given by the integration over the three-particle phase
space of the process $k_i+q\rightarrow k_j^{\mathrm{id.}}+k_k+k_l$, with the momentum fraction of particle $j$ fixed:
\begin{align}
\tilde{C}^{RR}_{p^{\prime}p}(\hx,\hz,Q^2,\epsilon)\propto\int \der \Phi_3(k_j,k_k,k_l;k_i,q)\delta \left(\hz-\hx\frac{(k_i+k_j)^2}{Q^2}\right)\left|(\Delta)\pazocal{M}^{RR}_{p^\prime p}\right|^2\,  .
\end{align}
They can be expressed as cuts of two-loop integrals in forward
kinematics, with $\hat{z}$ expressed as a linear cut propagator. Using integration-by-parts (IBP) identities we reduce these integrals to master integrals. The RR contributions to the SIDIS
coefficient functions are expressed in terms of 13 integral families, for a total of 21 master integrals. The master integrals are determined by
solving their differential equations in $\hat{x}$ and
$\hat{z}$. The boundary terms obtained by integrating the generic solutions over $\hat{z}$ and
comparing to the master integrals of the inclusive cases.  Their derivation is described in detail in Section 3 of~\cite{Bonino:2024adk}, where the antenna subtraction method for identified hadrons at NNLO was extended to processes with partonic initial states. 

The VV contributions correspond to the well-known two-loop quark form
factor~\cite{Gehrmann:2005pd} in space-like kinematics.

For the polarized coefficient functions, we use the same projectors as in the inclusive calculation of~\cite{Zijlstra:1993sh}. 
The external projectors contain two inherently four-dimensional objects, $\gamma_5$ and $\varepsilon^{\mu\nu\rho\sigma}$,  which must be treated consistently within dimensional regularization \cite{tHooft:1972tcz}. 
We use the Larin prescription~\cite{Larin:1991tj,Larin:1993tq}, which is derived from the 't Hooft-Veltman-Breitenlohner-Maison (\HVBM) scheme~\cite{tHooft:1972tcz,Breitenlohner:1977hr}, and consists of evaluating the Dirac traces in $d$ dimensions after setting
\begin{equation}
    \gamma_\mu \gamma_5 = \frac{i}{3!} \varepsilon_{\mu\nu\rho\sigma}\gamma^{\nu}\gamma^{\rho}\gamma^{\sigma} \, .
\end{equation}
The two remaining Levi-Civita tensors are contracted into $d$-dimensional metric tensors.

The virtual and double-virtual contributions to $g_1^h$ 
are described by the respective vector form factors \cite{Gehrmann:2005pd} rather than by their axial counterparts. Indeed
the photon couples to the quark line through a vector coupling, whereas the antisymmetric current carried by the photon is contracted only from the external leg. As a consequence, traces of quark-loops coupling to the polarized photon do not give rise to the axial anomaly \cite{Adler:1969gk,Bell:1969ts}.

The renormalization of $\alpha_s$ is carried out in the $\MSbar$ scheme to remove poles of 
ultraviolet origin. The leftover infrared poles are eliminated by mass factorization 
on the polarized and unpolarized PDFs and unpolarized FFs. 
In the polarized case, at this stage, all coefficient functions are still formulated in the Larin scheme. Consequently, the mass factorization 
counterterms of the polarized PDF are taken
in the Larin scheme as well, constructed from the 
polarized spacelike splitting functions in this scheme. 

Quantities in Larin and  $\MSbar$ are related by a finite scheme transformation~\cite{Zijlstra:1993sh,Matiounine:1998re,Mertig:1995ny,Vogelsang:1996im,Moch:2014sna}.

\section{Results}\label{sec:res}
In this section we report the analytical expressions for the $\MSbar$ unpolarized transverse and polarized pure-singlet (PS) coefficients of eq.~\eqref{CFNNLOlist}. The other coefficient functions can be found in  computer-readable ancillary files of the $\mathtt{arXiv}$ submission of \cite{Bonino:2024qbh,Bonino:2024wgg}. 

The polarized PS contribution is
{ \allowdisplaybreaks
\begin{align}
 \begin{autobreak}
\Delta \mathcal{C}^{\mathrm{PS}}_{qq} =
C_F
\Bigg\{
  s_{x/z} P_1(x,z)  \bigg(  - 4 \ln(s_{x/z}) \arctan(s_{x/z})
       + 4 \ln(z s_{x/z}) \arctan(z s_{x/z})
       + 2 \Ti2(s_{x/z})
       - 2 \Ti2(-s_{x/z})
       - 4 \Ti2(z s_{x/z})
       \bigg)

       + \delta(1-z)  \bigg( \bracket{x - 3} \ln(x)
       + 2 \bracket{x - 1}
       - \bracket{x + 2} \frac{1}{2} \ln(x)^2
      \bigg)

       - 8 \frac{x z - x - z + 1}{z}
       + 3 \frac{x z - x + z - 1}{z} \ln(x)
       + 3 \frac{x z + x - z - 1}{z} \ln(z)
       - 2 \frac{x z + x + z + 1}{z} \ln(x) \ln(z) \Bigg\} \, ,
       \end{autobreak}
\end{align}
}
while the transverse component of the  unpolarized PS contribution reads
{ \allowdisplaybreaks
\begin{align}
\begin{autobreak}
\mathcal{C}^{T,\mathrm{PS}}_{qq} =
C_F
\Bigg\{
  s_{x/z} P_2(x,z)  \bigg(  - \frac{1}{8} \ln(s_{x/z}) \arctan(s_{x/z})
       + \frac{1}{8} \ln(z s_{x/z}) \arctan(z s_{x/z})
       - \frac{1}{8} \Ti2(z s_{x/z})
       + \frac{1}{16} \Ti2(s_{x/z})
       - \frac{1}{16} \Ti2(-s_{x/z})
       \bigg)

       - 2 \frac{x z + x + z + 1}{z} \ln(x) \ln(z)
       - \frac{1}{16} \ln(x) P_3(x,z)
       + \frac{1}{16} \ln(z) P_4(x,z)
       - \frac{5}{8} P_5(x,z) \Bigg\}

 \, ,
       \end{autobreak}
   \end{align}
   }
   with $s_{x/z}=\sqrt{\frac{x}{z}}$, the inverse tangent integral function defined by
   \begin{equation}
   \mathrm{Ti_2}(y)=\int_0^y\frac{\arctan x}{x} \mathrm{d} x \,  ,
   \end{equation}
   and rational functions
   \begin{align}
   P_1(x,z) &=
  \frac{x^2 z + x z^2 + x + z}{x z}\, , \nonumber \\
P_2(x,z) &=
  \frac{5 x^4 z^2 + 18 x^3 z^3 + 18 x^3 z + 5 x^2 z^4 + 52 x^2 z^2 + 5 x^2 + 18 x z^3 + 18 x z + 5 z^2}{x^2 z^2} \, , \nonumber \\
  P_3(x,z) &=
  \frac{5 x^3 z^2 - 5 x^3 z - 5 x^2 z^3 - 34 x^2 z^2 + 34 x^2 z + 5 x^2 - 5 x z^3 - 34 x z^2 + 34 x z + 5 x + 5 z^2 - 5 z}{x z^2} \, , \nonumber \\
P_4(x,z) &=
  \frac{5 x^3 z^2 + 5 x^3 z - 5 x^2 z^3 + 34 x^2 z^2 + 34 x^2 z - 5 x^2 + 5 x z^3 - 34 x z^2 - 34 x z + 5 x - 5 z^2 - 5 z}{x z^2}\, , \nonumber \\
  P_5(x,z) &=
  \frac{x^3 z^2 - x^3 z + x^2 z^3 + 6 x^2 z^2 - 6 x^2 z - x^2 - x z^3 - 6 x z^2 + 6 x z + x - z^2 + z}{x z^2}\, .

   \end{align}
   The pure-singlet contributions arise from diagrams in which the quark line that couples to the photon goes entirely unresolved and is connected to the initial state parton via a gluon.  Since the quark-antiquark pair is resolved by the photon and the other quark line is constrained by the initial and final-state kinematics, no contribution propotional to $\delta(1-z)$ is expected in this channel. Nevertheless, in the polarized case, a delta distribution appears from the pure-single contribution of Larin to $\overline{\mathrm{MS}}$ scheme transformation.
   
\begin{figure}[t]
  \includegraphics[width=\textwidth]{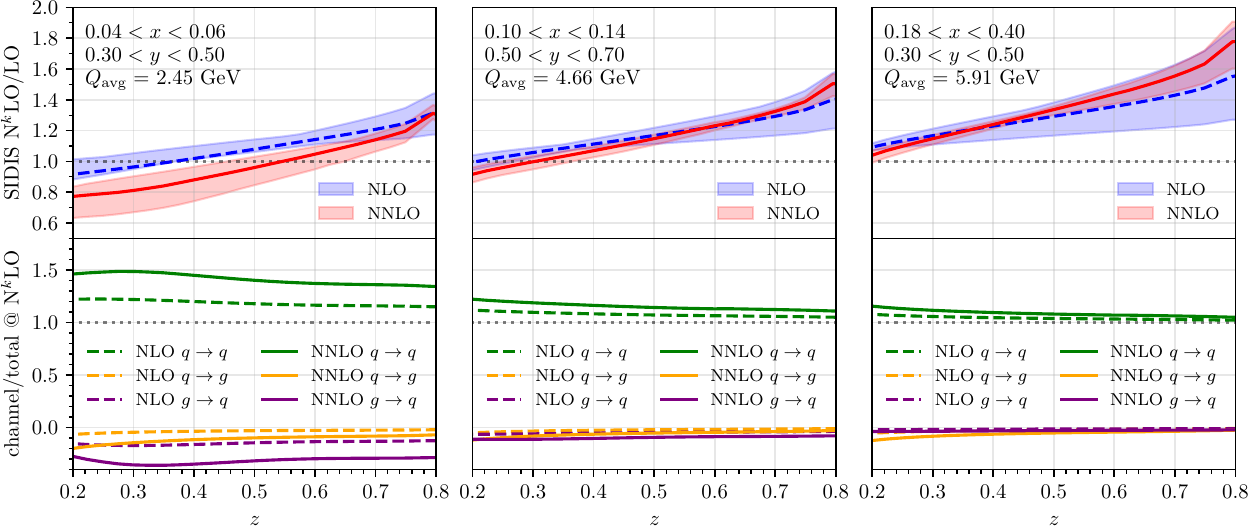}
  \caption{ QCD $K$-factors up to NNLO and fractional contribution of individual channels. The channels that are not shown are numerically negligible.  To determine the scale variation band we vary by factors of 2 around the central scale ($\mu_R=\mu_F=\mu_A=Q$) fixing $\mu_F=\mu_A$.}
  \label{fig:kfact}
\end{figure}

 The numerical impact of the QCD corrections to eq.~\eqref{d3sigdxdydz} is illustrated in the form of K-factor for a few selected bins of the COMPASS single-inclusive pion production measurement~\cite{COMPASS:2016xvm} in Figure~\ref{fig:kfact}.  We use the NNPDF3.1 PDF set~\cite{NNPDF:2017mvq} and the FF set from~\cite{Borsa:2022vvp} at NNLO, with $\alpha_s(M_Z)=0.118$ and with $N_F = 5$ light quarks. 
   
Our polarized and unpolarized coefficient functions are in agreement for all channels with the results presented in \cite{Goyal:2023zdi,Goyal:2024tmo}.

 \section{Conclusions}
 The full set of NNLO corrections to the SIDIS coefficient functions is now available, for both polarized and unpolarized scattering. In these proceedings we reviewed the analytical derivation and showed selected analytical expressions.  The unpolarized results will enable the extraction of FFs from global fits at NNLO accuracy.  Furthermore,  polarized SIDIS is a key observable for accessing the polarized flavour structure of the proton and will be a key ingredient for the BNL Electron-Ion Collider (EIC) spin program~\cite{Aschenauer:2019kzf}. 

\bibliographystyle{JHEP}
\bibliography{bonino}

\end{document}